\documentclass[letter]{aa} 

\usepackage{graphicx}
\usepackage{txfonts}

\newcommand{\hi}{\mbox{H{\sc i}}}

\newcommand{\kms}{km s$^{-1}$}

\newcommand{\msol}{\rm M$_\odot$}
\newcommand{\lsol}{$L_\odot$}
 
\begin{document}
 
 \title{\hi\ Observations of two New Dwarf Galaxies: Pisces A \& B\\
with the SKA Pathfinder KAT-7}
 
 \titlerunning{KAT-7 Observations of Pisces A \& B}
 \authorrunning{Carignan et al.} 
 
 \author{C. Carignan\inst{\ref{inst1}}\fnmsep \inst{\ref{inst2}} \thanks{ccarignan@ast.uct.ac.za}\and Y. Libert\inst{\ref{inst1}}\and D. M. Lucero\inst{\ref{inst1}} \fnmsep \inst{\ref{inst3}}\and T. H. Randriamampandry\inst{\ref{inst1}}\and T. H. Jarrett\inst{\ref{inst1}}\and \\
T. A. Oosterloo\inst{\ref{inst3}} \fnmsep \inst{\ref{inst4}}\and E. J. Tollerud\inst{\ref{inst5}}}
          
  \institute{
  Department of Astronomy, University of Cape Town, Private Bag X3, Rondebosch 7701, South Africa\label{inst1}
  \and
  Observatoire d'Astrophysique de l'Universit\'e de Ouagadougou, BP  7021, Ouagadougou 03, Burkina Faso\label{inst2}
  \and
  Netherlands Institute for Radio Astronomy (ASTRON), Postbus 2, 7990 AA Dwingeloo, The Netherlands\label{inst3}
  \and
  Kapteyn Astronomical Institute, University of Groningen, PO Box 800, 9700 AV Groningen, The Netherlands\label{inst4}
  \and
  Space Telescope Science Institute, 3700 San Martin Dr, Baltimore, MD 21218, USA\label{inst5}
               } 

\abstract
{Pisces A \& Pisces B are the only two galaxies found via optical imaging and spectroscopy out of 22 \hi\ clouds identified in the GALFAHI survey as dwarf galaxy candidates.}
{Derive the \hi\ content and kinematics of Pisces A \& B.}
{Aperture synthesis \hi\ observations using the seven dish Karoo Array Telescope (KAT-7), which is a pathfinder instrument for MeerKAT, the South African precursor to the mid-frequency Square Kilometre Array (SKA-MID).}
{The small rotation velocities detected of $\sim$5 \kms\ and $\sim$10 \kms\ in Pisces A \& B respectively, and their \hi\ content show that they are really dwarf irregular galaxies (dIrr). Despite that small rotation component, it is more the random motions $\sim$ 9-11 \kms\ that provide most of the gravitational support, especially in the outer parts. The study of their kinematics, especially the strong gradients of random motions, suggest that those two dwarf galaxies are not yet in equilibrium.}
{These \hi\ rich galaxies may be indicative of a large population of dwarfs at the limit of detectability. However, such gas-rich dwarf galaxies will most likely never be within the virial radius of MW-type galaxies and become sub-halo candidates. Systems such as Pisces A \& B are more likely to be found at a few Mpc.s from MW-type galaxies. 
}


\maketitle
%

\section{Introduction}

Optical searches for faint dwarf galaxies and satellites  (dSph or dIrr) are generally surface brightness limited. While having some success recently for very nearby dwarfs  in deep optical surveys, such as the Dark Energy Survey \citep{kop15} and the PAN-STARRS 1 3$\pi$ Survey  \citep{lae15}, the task becomes very difficult as soon as one gets further out in the Local Volume (5-10 Mpc). This was shown clearly with the discovery of seven dwarf galaxies close to M101, which have central surface brightnesses of $\mu_g \sim 25.5-27.5$ mag arcsec$^{-2}$, well below the sky brightness, and was only possible through the development of  a new instrument, the Dragonfly Telephoto Array \citep{mer14}.

This motivates searches for dwarf galaxies using the 21cm emission line of neutral hydrogen (\hi). While such searches cannot identify passive dwarf galaxies (dSph) like most Local Group satellites, which lack \hi\ \citep[M$_{\small \hi}$/L$_V$ $\leq$ 10$^{-3}$ \msol/\lsol:][]{spe14}, they have the potential to find gas-rich, potentially star-forming dwarf (dIrr) galaxies  \citep[see e.g.][]{ada13, ada15, can15, jan15}. 

Good examples are the dwarf galaxies Leo P, found first in \hi\  \citep{gio13} and later confirmed by optical imaging \citep{rho13} and Leo T, first detected in \hi\ by HIPASS \citep{won06} but  discovered independently in the optical a year later by \citet{irw07}. However, those galaxies detected in \hi\ cannot aleviate the Missing Satellite Problem and be sub-halos candidates, since, for example, Leo T at $\sim$400 kpc \citep{mcc12} is outside the virial radius of the Milky Way (MW) and Leo P at $\sim$ 1.5-2.0 Mpc \citep{mcq13} is well outside the Local Group. Possible sub-halos, that could be associated to the MW or M31,  are more likely dSphs stripped of their gas \citep{grc09} or completely "dark" systems \citep{bro13}.

\citet{tol15}, hereafter TOL15, report the discovery of two dwarf galaxies, Pisces A and B, from a blind 21cm \hi\ search. These were the only two galaxies found via optical imaging and spectroscopy (WIYN) out of 22 \hi\ clouds identified in the GALFAHI survey \citep{pee11} as dwarf galaxy candidates. 
While the distance uncertainty made any interpretation ambiguous at the time, TOL15 propose that they are likely winthin the Local Volume (< 10 Mpc) but outside of the Local Group (> 1 Mpc). They also suggest that they may be among the faintest star-forming dwarf galaxies known. 

\begin{figure}
\hspace{-25pt}
\begin{tabular}{ll}
\includegraphics[angle=-90,width=57mm]{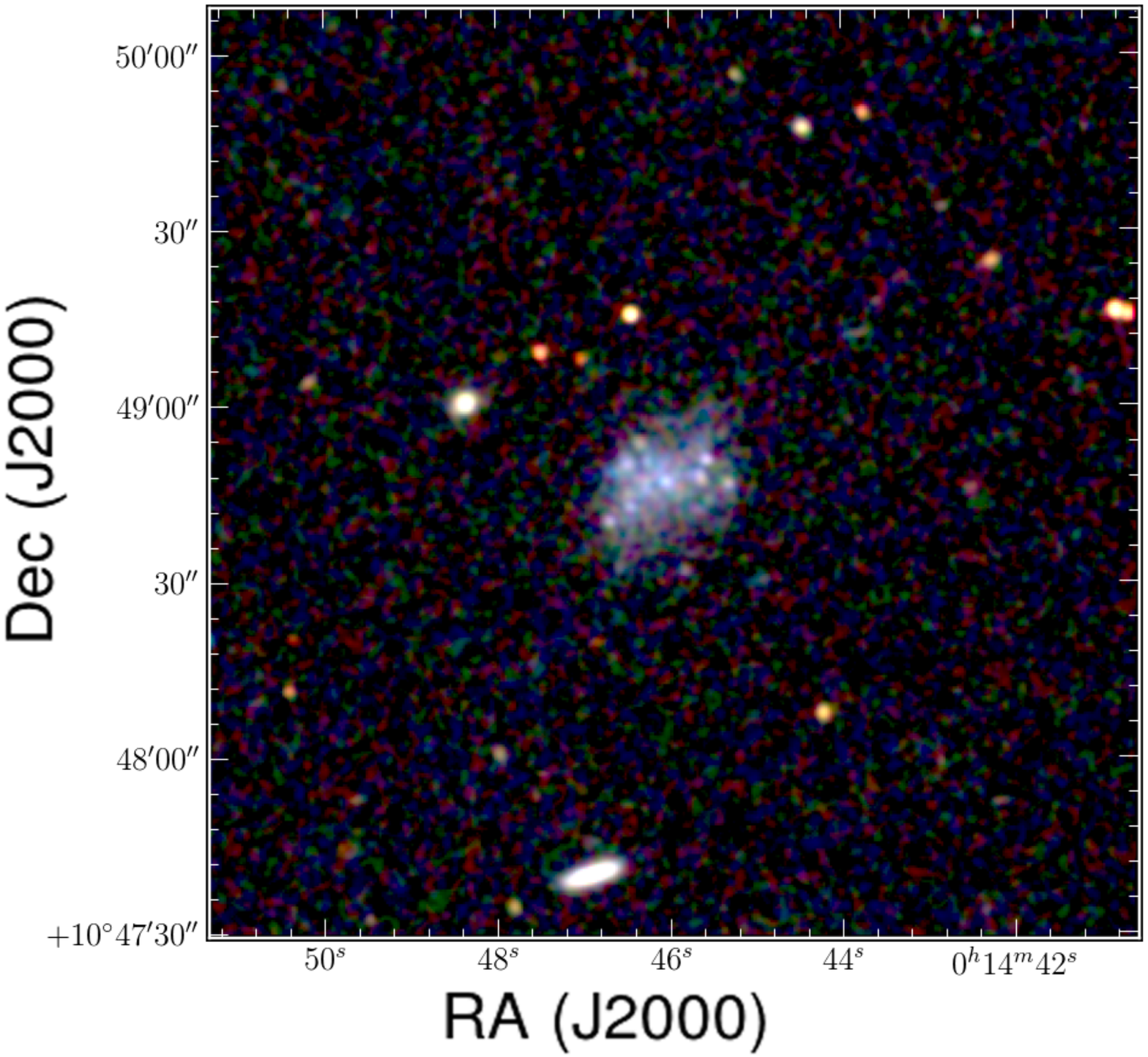}
\hspace{-35pt}
\includegraphics[angle=-90,width=57mm]{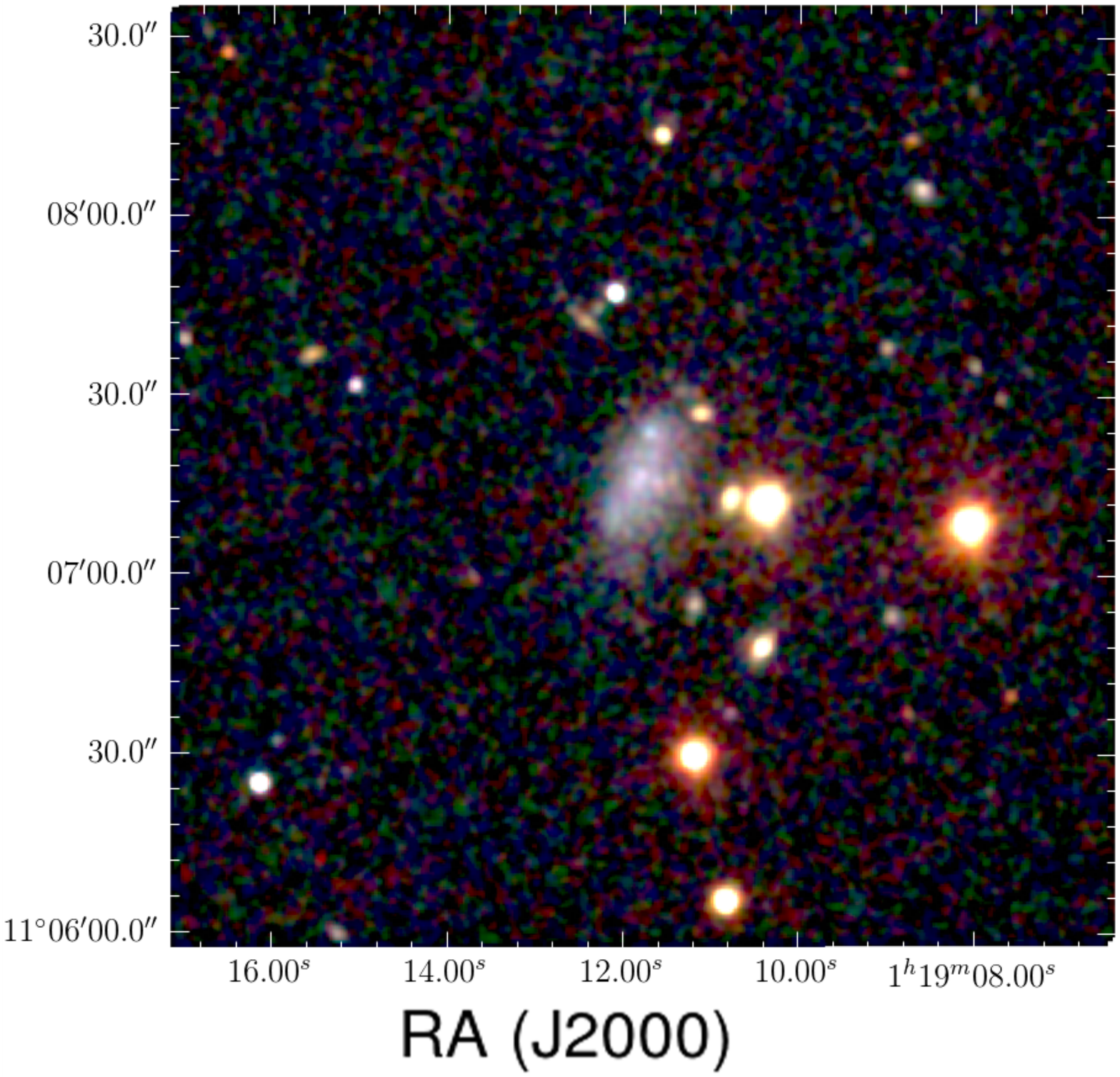}
\end{tabular}
\caption{SDSS combined gri images of Pisces A (left) \& B (right).}
\label{fig:images}
\end{figure}

Fig. \ref{fig:images} shows SDSS (gri)  images of Pisces A and B. The fact that the brightest stars are being resolved in Pisces A already suggests that it is most likely closer than Pisces B.
Table \ref{para} summarizes the parameters of the two new dwarf galaxies, using the new determined distances of \citet{tol16}.
The remainder of this letter is as follows. Sect.~\ref{sec:ODR} gives a description of the Karoo Array Telescope (KAT-7) observations and data reduction,  Sect.~\ref{sec:HICP} describes the \hi\ content and kinematics of Pisces A and B, the main results are discussed in Sect.~\ref{sec:dis} and Sect.~\ref{sec:con} presents the final conclusion.

\begin{table}
\caption{Parameters of Pisces A and B.}
\label{para}
\centering
\begin{tabular}{lrr}
\hline\hline
Parameter  & Pisces A & Pisces B\\        
\hline
R. A.  (J2000)                        &00$^{\rm h}$ 14$^{\rm m}$ 46.0$^{\rm s}$ &01$^{\rm h}$ 19$^{\rm m}$ 11.7$^{\rm s}$      \\
Dec.  (J2000)                                &+10$^{\rm o}$  48\arcmin\ 47\arcsec&+11$^{\rm o}$  07\arcmin\ 18\arcsec     \\
$l$ ($^{\rm o}$)                                                	    &108.52	& 133.83                            \\
$b$ ($^{\rm o}$)                                                      & $-51.03 $ 	& $-51.16$                            \\
Axis ratio (q=b/a) 					&0.60 & 0.45 \\
Inclination ($^{\rm o}$)					& 53$\pm$5 & 63$\pm$5 \\
P. A. ($^{\rm o}$)						& 111$\pm$3 & 156$\pm$1 \\
NUV (mag)                                                  & 19.18$\pm$0.13 & 18.80$\pm$0.06 \\
$m_r$ (mag)                                                & 17.37$\pm$0.19 & 17.19$\pm$0.16 \\
($g - r$) (mag)                              		&  0.24 & 0.31\\
Distance (Mpc)$^{\rm a}$						& 5.6$\pm$0.2  & 8.9$\pm$0.8 \\ 
Scale (kpc/arcmin)            				& 1.6 		& 2.6	 \\
$r_{\rm eff , major}$ (pc)$^{\rm a}$                                    & $145^{+5}_{-6}$     & $323^{+27}_{-30}$  \\			
$M_V$ (abs. mag)$^{\rm a}$				&	$-11.57^{+0.06}_ {-0.05}$		&  $-12.90^{+0.2}_ {-0.2}$					\\
log ($M_{*,\ SFH}$ / $M_{\odot}$)$^{\rm a}$                           & 7.0$^{+0.4}_ {-1.7}$                  &   7.5$^{+0.3}_ {-1.8}$         \\
\hline
(a) \citet{tol16}
\end{tabular}
\end{table}

\section{KAT-7 Observations and Data Reduction}
\label{sec:ODR}

The observations were obtained with the seven-dish KAT-7 array \citep{car13, luc15}, located close to the South African SKA core site in the Northern Cape's Karoo desert region. The array is extremely compact, with baselines ranging from 26 m to 185 m and receivers having a very low $T_{sys} \sim 26$ K, which makes it very sensitive to large scale, low surface brightness emission. The parameters of the KAT-7 observations are given in Table~\ref{kat7par}.
The data were collected between 2014 November 27 and 2014 December 16 in the c16n13M4K spectral line mode.  This correlator mode gives velocity channels of 0.64 km s$^{-1}$ over a flat bandpass of $\sim$2000 km s$^{-1}$. When producing the final cubes, we averaged 5 channels for a final channel width of 3.2 \kms.

Each of the observing sessions, of typically 6 hours, were reduced separately.  All data calibration was done using standard calibration tasks in the Common Astronomy Software Applications (CASA 4.2.0) package \citep{mcm07}. Phase drifts as a function of time were corrected by means of a nearby point source (0022+002) observed every 30 minutes.  This source was also used to correct for variations in the gain as a function of frequency (bandpass calibration).  The absolute flux scale was set by observing 1934-638.  Comparisons of the flux measurements on the observed calibrators suggest that the absolute flux uncertainties are on the order of 5\%.  Variations in the bandpass are on the order of 1\%.  

Continuum emission was subtracted from the raw UV data by making first order fits to the line free channels using the CASA task  {\sc uvcontsub}.  The calibration was then applied and Pisces A and B were {\sc split} from the calibration sources. KAT-7 does not use doppler tracking and CASA does not fully recognize frequency keywords, so special care was taken to produce uv data sets and test cubes with the proper velocity coordinates \citep[see][]{car13}. The individual calibrated continuum subtracted UV data sets were then combined together using the CASA task {\sc concat}.  

Preliminary imaging of the combined data in CASA revealed the presence of artifacts in the form of horizontal lines, caused by some unidentified internal instrumental feature. At the suggestion of T. A. Oosterloo, these were removed by flagging all visibilities near u = 0 \citep[see also][]{hes15}, clearly seen to be out of range after performing an FFT on the images. 
Careful data analysis also allowed us to discover and correct for an artifact caused by {\it crosstalk} between two antennas. 
This occurs when one antenna is pointing towards the back of another at an elevation high enough to prevent the shadowing flag to be triggered by the system, but low enough to produce strong variations in the bandpass.

\begin{table}
\caption{Parameters of the KAT--7  observations.}
\label{kat7par}
\begin{tabular}{lr}
\hline\hline
Parameter & Pisces A \& B  \\
\hline
Start of observations &27 November 2014 \\
End of observations &16 December 2014 \\
Total integration &  26.5 \& 24.0 hours\\
FWHM of primary beam & 1.27\degr \\
Total Bandwidth & 12.5 MHz \\
Central frequency &  1419.3 \& 1417.5 MHz \\
Channel Bandwidth & 15.0 kHz \\
Channel width & 3.2 \kms \\
Maps gridding & 15\arcsec\ x 15\arcsec \\
Maps size & 512 x 512 \\
Flux calibrator	& 1934-638 \\
Phase/bandpass calibrator  & 0022+002 \\
\hline
Robust = 0 weighting function & \\
FWHM of synthesized beam & 283\arcsec x 184\arcsec  \& 284\arcsec x 181\arcsec \\
RMS noise (mJy/beam)  & 4.5 \& 4.2  \\
Column density limit & \\
(3$\sigma$ over 16 \kms)  &   $\sim 5 \times 10^{18}$ cm$^{-2}$ \\
\hline
Natural weighting function & \\
FWHM of synthesized beam & 320\arcsec x 203\arcsec  \& 316\arcsec x 200\arcsec \\
RMS noise (mJy/beam )  & 3.4 \& 3.2 \\
Column density limit   & \\
(3$\sigma$ over 16 \kms)  & $\sim 2.5 \times 10^{18}$ cm$^{-2}$ \\
\hline
\end{tabular}
\end{table}


\section{$\hi$ Content and Kinematics}
\label{sec:HICP}

\begin{figure}
\includegraphics[width=90mm]{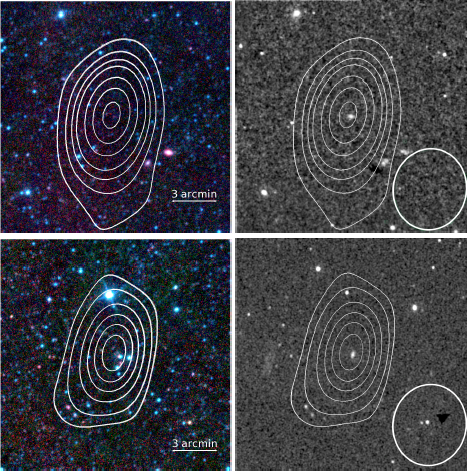}
\caption{\hi\ distributions in Pisces A (top) and Pisces B (bottom), superposed on the 3-color WISE w1+w2+w3 composite (left) and Galex NUV (right) images, from the natural weighted cubes.
Contours are at 0.3 (3 $\sigma$), 0.6, 1.2, 1.8, 2.4 and 3.0 $\times 10^{19}$ cm$^{-2}$.}
\label{fig:mom0}
\end{figure}

The total \hi\ distribution maps of Pisces A \& B, shown in Figure \ref{fig:mom0}, were derived using the task {\sc sqash} in AIPS \citep{gre03}. They are superposed on the WISE (w1+w2+w3) and Galex (NUV) images. As expected, not much IR flux is seen in the WISE images due to the absence of an appreciable old stellar population
(Pisces A is not detected in WISE, but Pisces B is, which suggests it has a larger stellar mass)  
but the two star-forming dwarfs are clearly seen in the UV. The faintest level goes down to $\sim$3 $\times$ 10$^{18}$ cm$^{-2}$. At that level, the galaxies have diameters of $\sim$9.6\arcmin\ and $\sim$8.5\arcmin , respectively. Despite the large synthesized beam (see Table \ref{kat7par}), due to the northern declinations of the sources, there are still $\sim$2-3 beams across the objects, which is sufficient to detect any significant velocity gradient.

\begin{figure}
\centering
\vspace{-10pt}
\begin{tabular}{l}
\includegraphics[width=65mm]{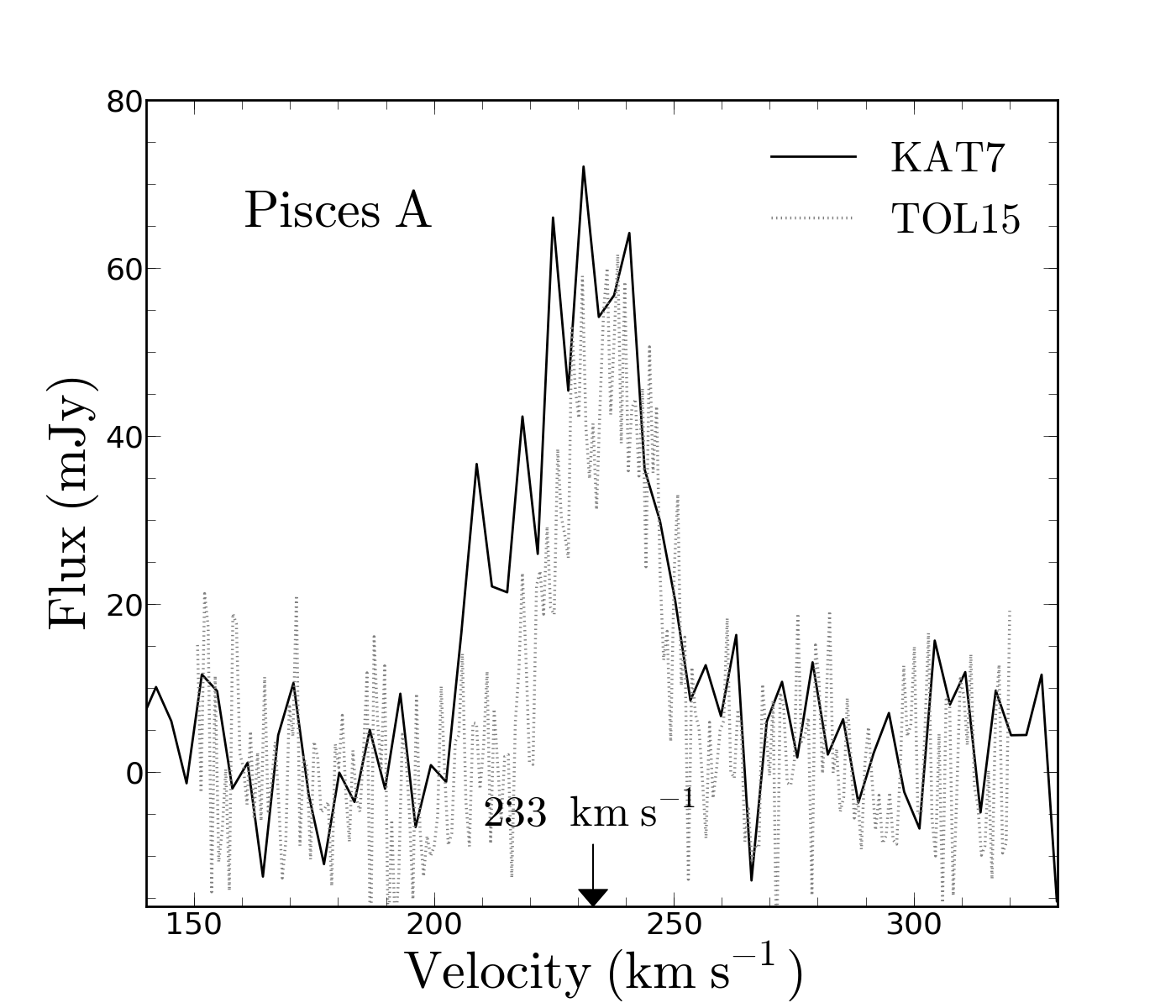}\\
\includegraphics[width=65mm]{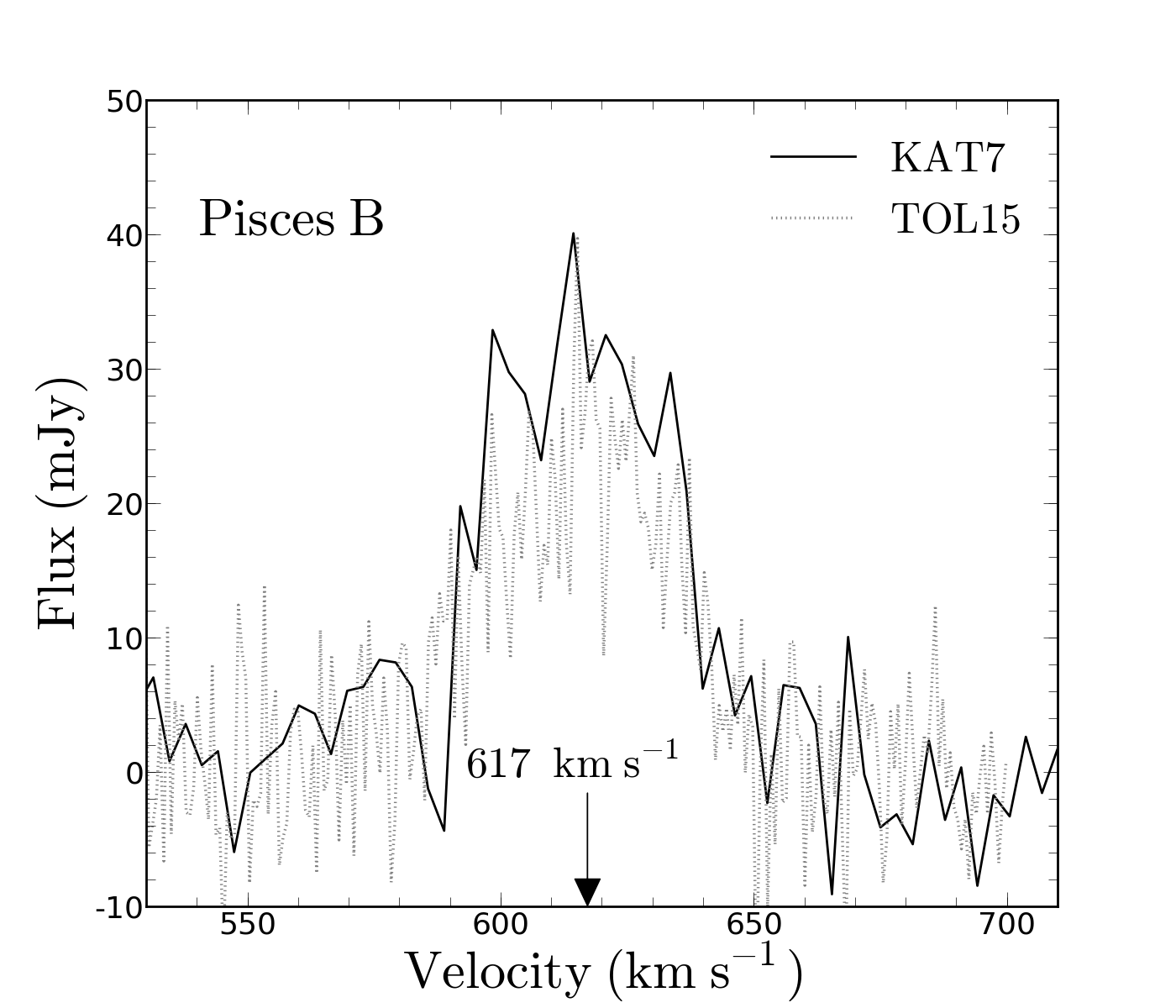}
\end{tabular}
\caption{Global \hi\ profiles of Pisces A (top) and Pisces B (bottom), using the natural weighted cubes. Velocities are heliocentric. The grey profiles are those from TOL15.}
\label{fig:global}
\end{figure}

The global \hi\ profiles are given in Figure \ref{fig:global}. They were obtained using the task {\sc blsum} in AIPS. Mid-point velocities (50\% level) of 233$\pm 3$ \kms\  and 617$\pm 3$ \kms\ are found for Pisces A and Pisces B, respectively. These can be compared to 236$\pm 0.5$ \kms\ and 615$\pm 1$ \kms\ found by TOL15. The profile widths at the 50\% levels are $\Delta$V$_{A} = 28 \pm 3$ \kms\ and $\Delta$V$_{B} = 41 \pm 3$ \kms, compared to $\Delta$V$_{A} = 22.5 \pm 1.3$ \kms\ and $\Delta$V$_{B} = 43 \pm 3$ \kms\ for TOL15. Total \hi\ fluxes of $1.68 \pm 0.20$ and $1.76 \pm 0.05$ Jy \kms\ are found, which correspond to \hi\ masses of $4.0 \pm 0.5$ $\times 10^{5}$ M$_{\odot}$ and  $4.2 \pm 0.2$ $\times 10^{5}$ M$_{\odot}$ at a fiducial distance of 1.0 Mpc. This is slightly larger than the values of $2.8 \pm 0.2$ $\times 10^{5}$ M$_{\odot}$ and  $3.8 \pm 0.4$ $\times 10^{5}$ M$_{\odot}$ found by TOL15. At the adopted distance, this corresponds to total \hi\ masses of $1.3\pm0.4 \times 10^7$  \& $3.3\pm1.0 \times 10^7$ \msol\ for Pisces A \& B.

\begin{figure*}
\centering
\vspace{-10pt}
\begin{tabular}{ccc}
\includegraphics[width=50mm]{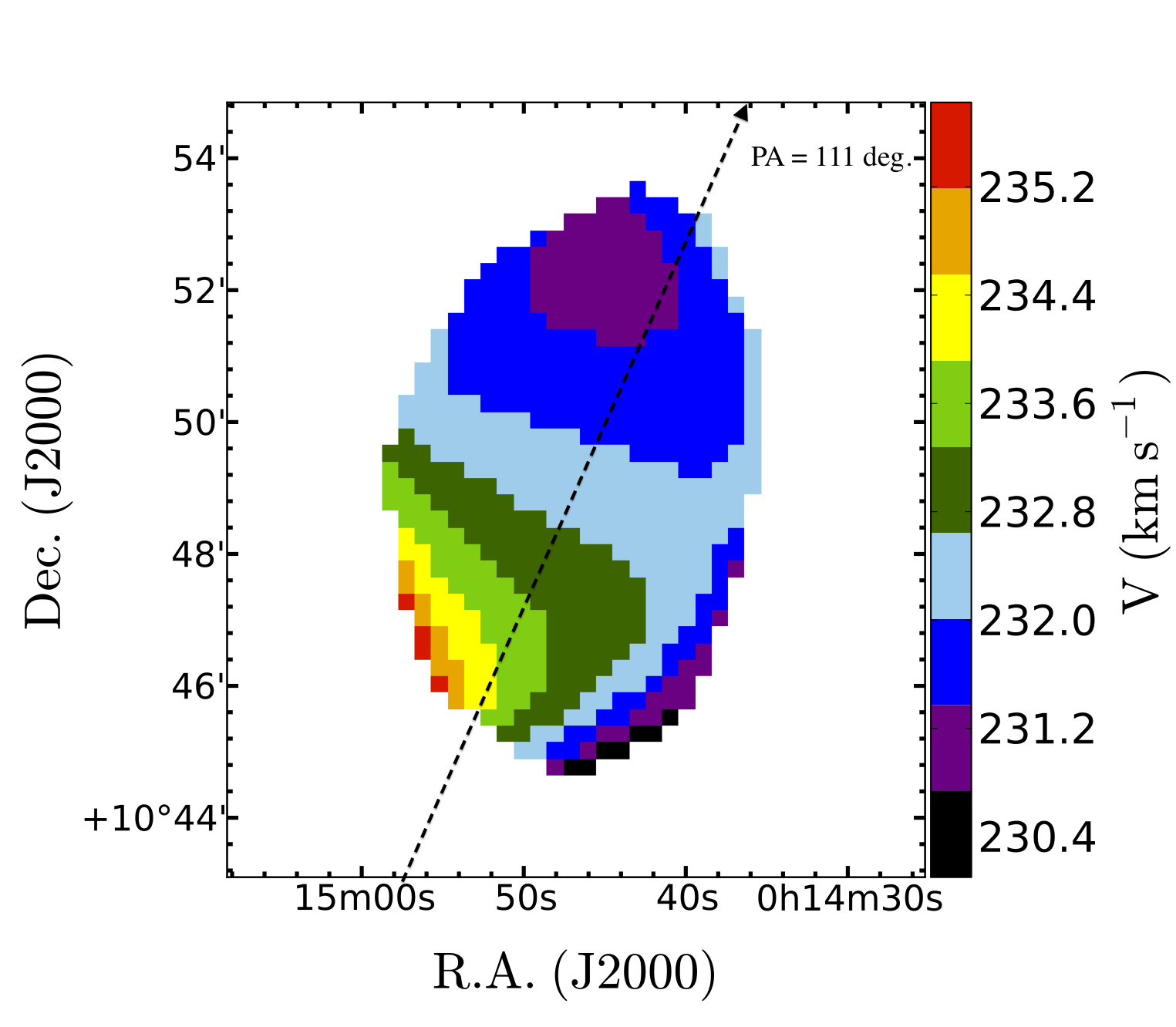}
\includegraphics[width=50mm]{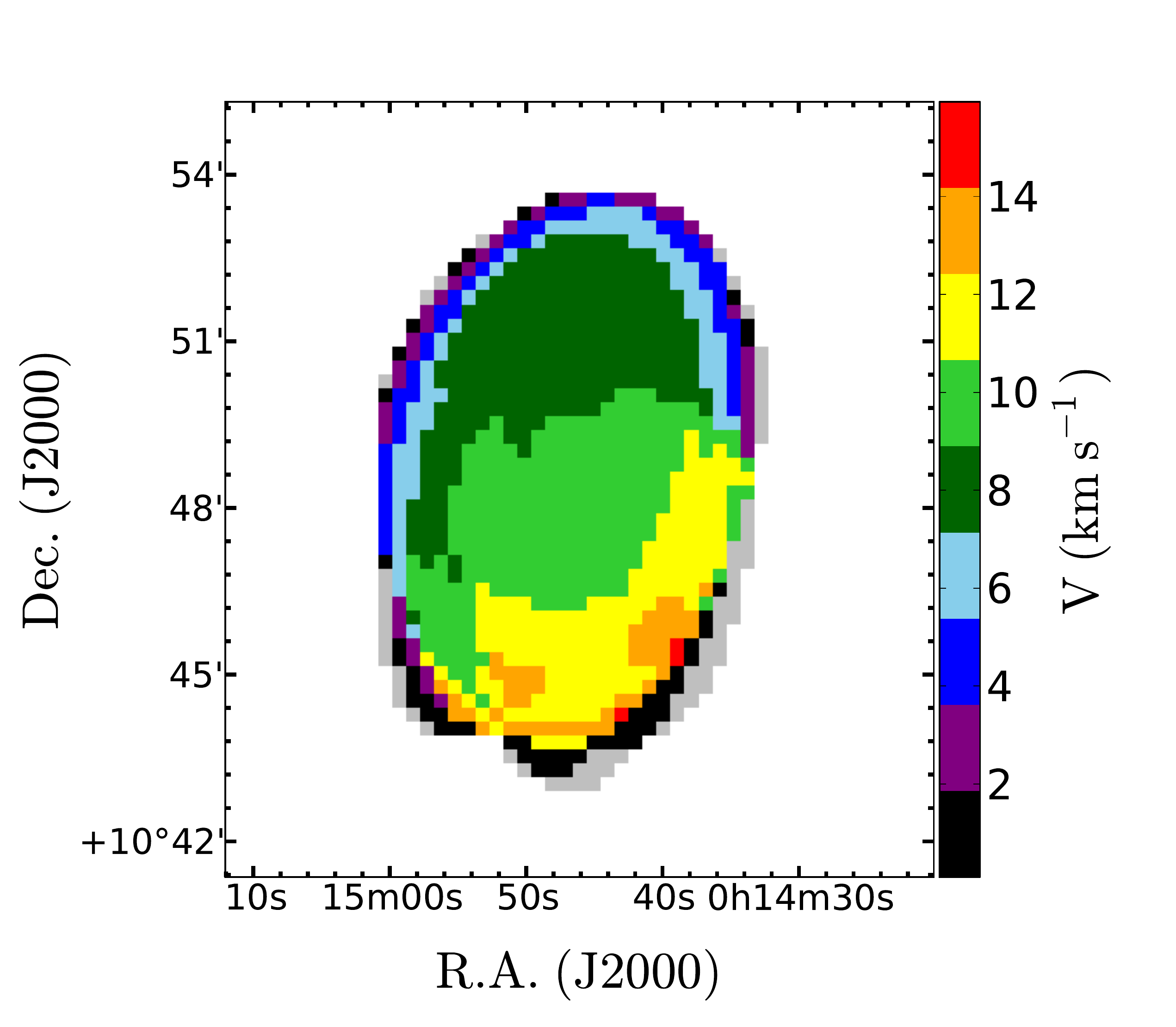}
\includegraphics[width=60mm]{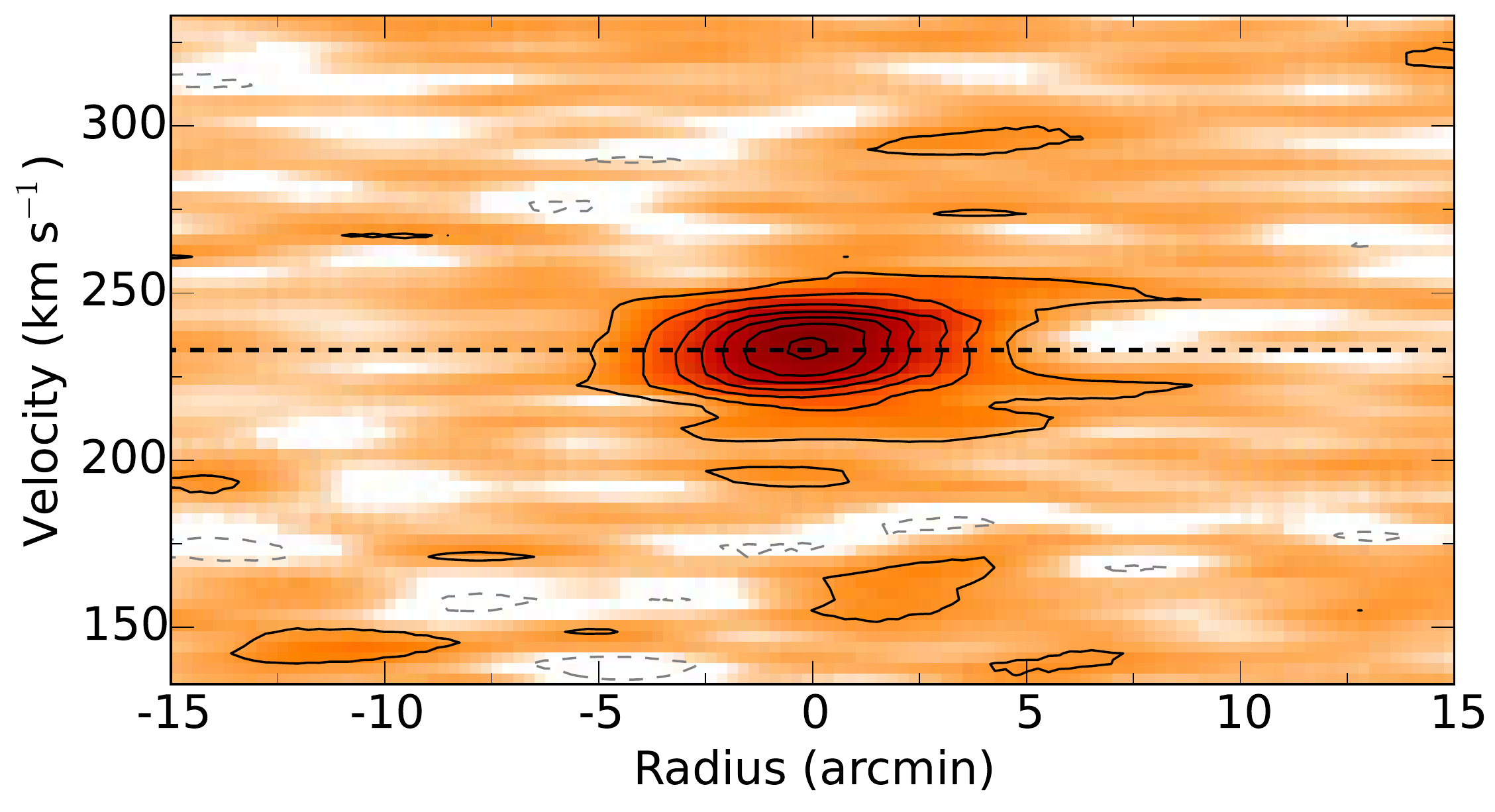}\\
\includegraphics[width=50mm]{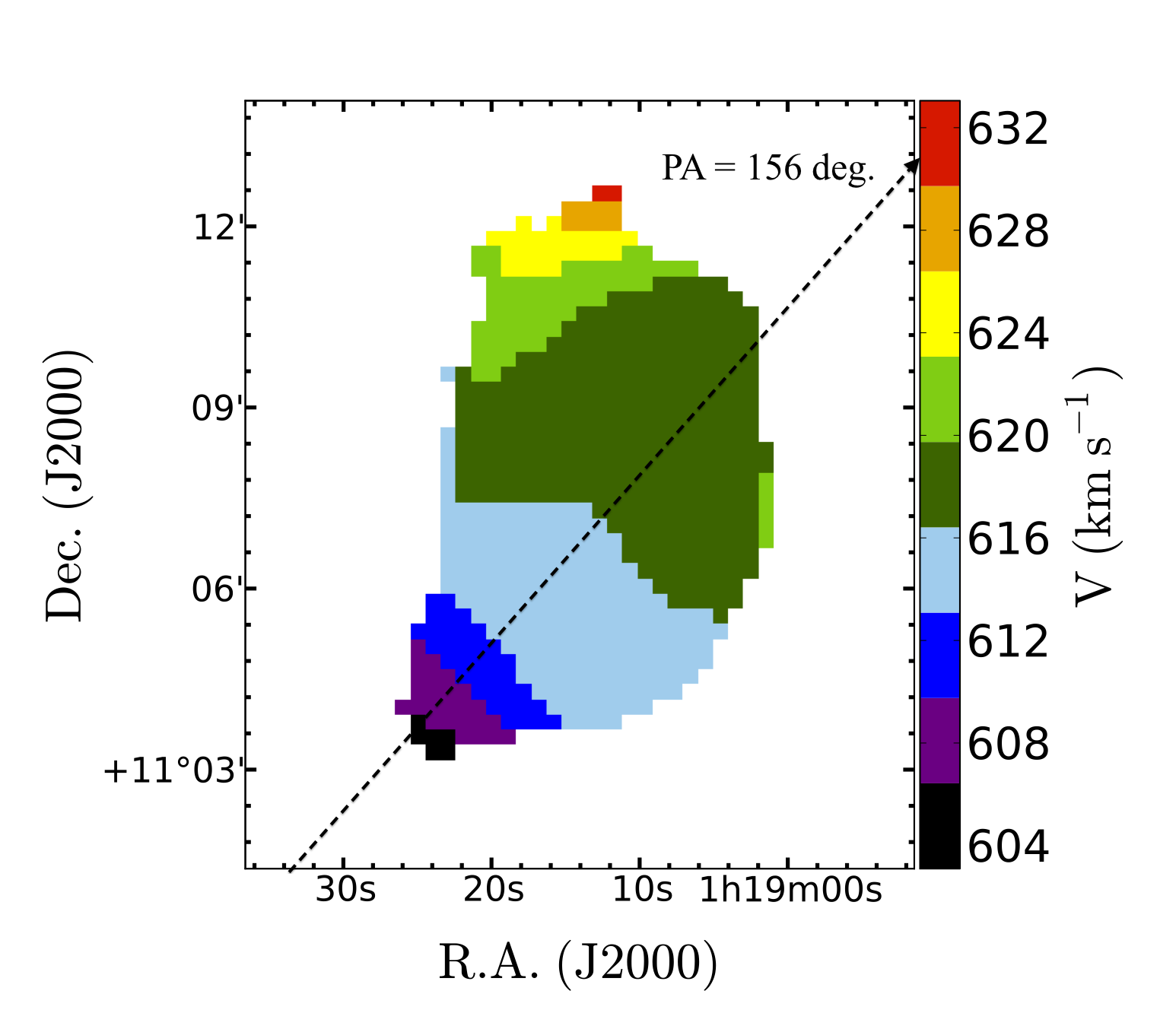}
\includegraphics[width=50mm]{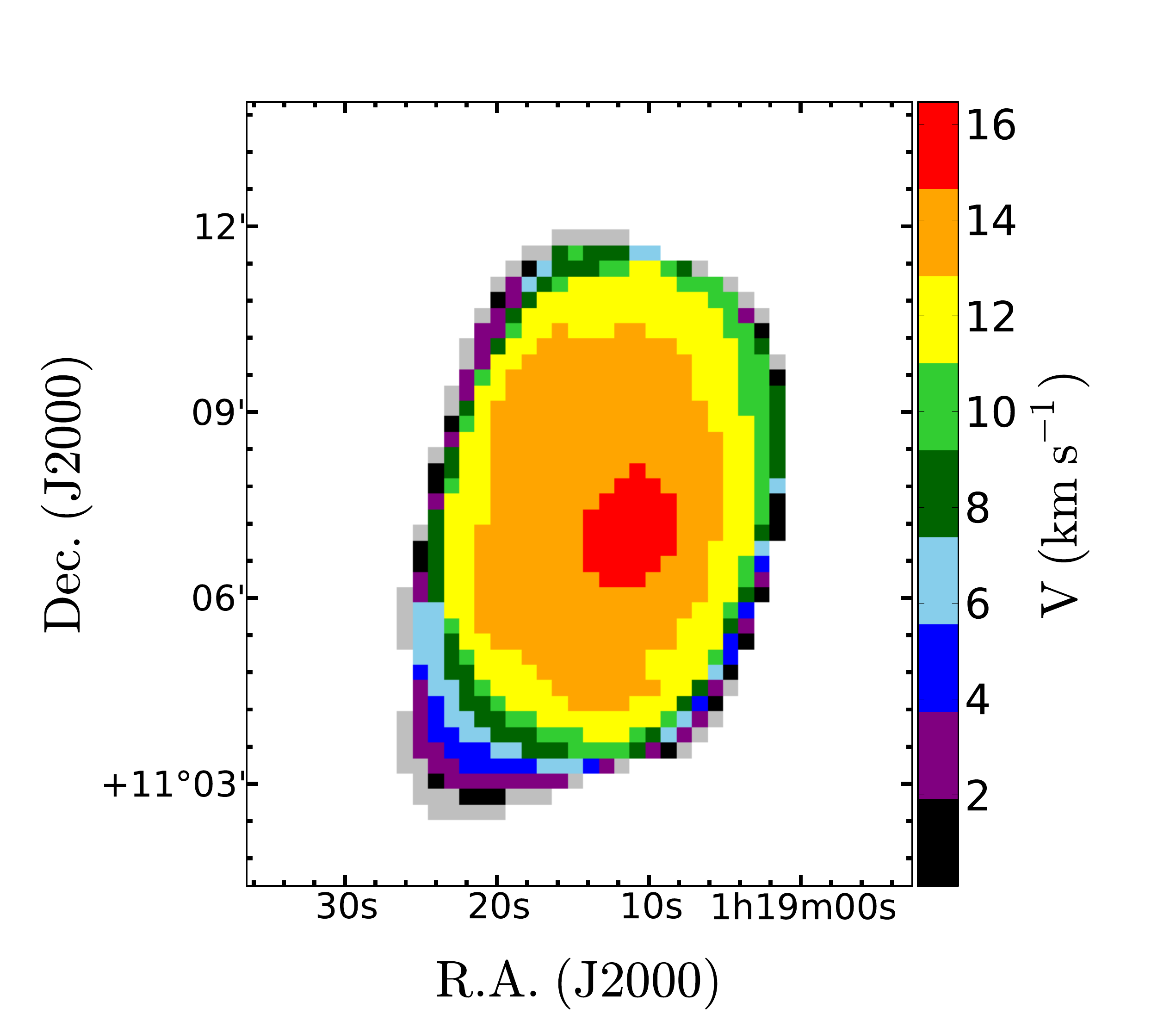}
\includegraphics[width=60mm]{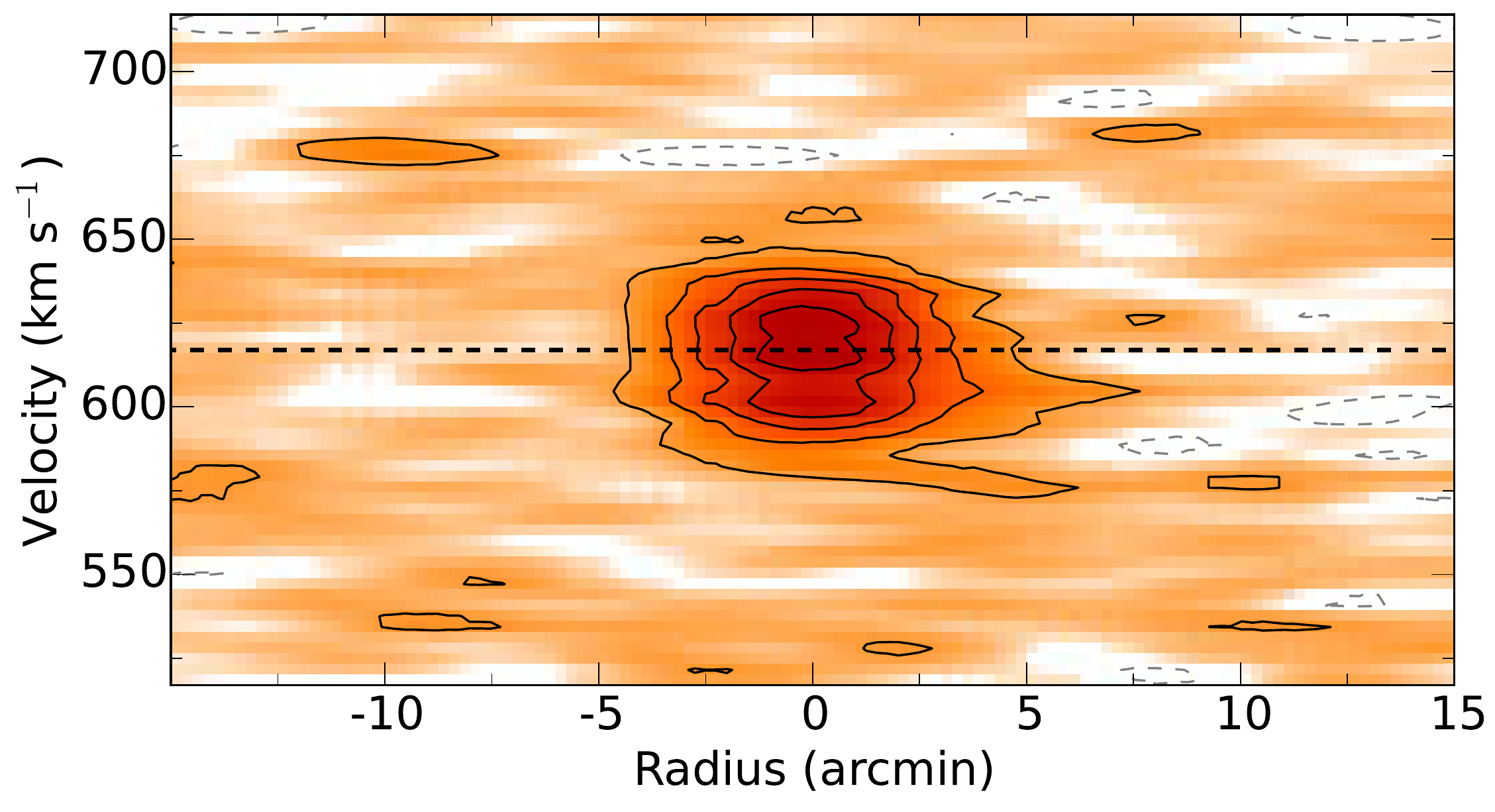}\\
\end{tabular}
\caption{Velocity fields (the dashed line shows the optical PA) from Gaussian profiles fitting (left), velocity dispersion from moment analysis (center) and PV diagrams of the hanning smoothed data for Pisces A (top) and Pisces B (bottom), where the dashed lines show the systemic velocities.} 
\label{fig:VF}
\end{figure*}

It is difficult to understand the difference for Pisces A ($\sim$30\% in flux and $\sim$20\% in width), while both data sets agree perfectly in the case of Pisces B, since the same calibrators and techniques were used for both objects. Comparing the global profiles, the difference is clearly on the approaching side. However, we are quite confident in the width derived for Pisces A since the value obtained by TOL15 would imply that no rotation is present, while rotation is clearly  seen in the velocity field of Pisces A (Fig. \ref{fig:VF}).

From the moment analysis, derived with the task {\sc momnt} in AIPS, the mean $\sigma$ (random motion) of the \hi\ is $9.4 \pm 2.0$ \kms\ for Pisces A and 
$11.0 \pm 3.7$ \kms\ for Pisces B. For Pisces A, there is a gradient from $\sim$6 \kms\ in the North to $\sim$12 \kms\ in the South and for Pisces B from $\sim$15 \kms\ in the center down to $\sim$7 \kms\ at the edge of the disk. 
The global profiles show that some rotation is clearly present of the order of ($7 \pm 3$)/2(sin$\it i$)$\sim4.6 \pm 2.0$ \kms\ and ($17 \pm 3$)/2(sin$\it i$)$\sim9.4 \pm 2.1$ \kms\ for Pisces A \& B. This is exactly what is seen in the velocity fields of Fig. \ref{fig:VF}, obtained by Gaussian profile fitting using the task {\sc xgaus} in AIPS. The derived \hi\ parameters are given in Table \ref{parahi}.

\begin{table}
\caption{\hi\ parameters of Pisces A and B.}
\label{parahi}
\centering
\begin{tabular}{lrr}
\hline\hline
Parameter  & Pisces A & Pisces B\\          
\hline
F$_{\rm HI}$ (Jy \kms) & $1.7 \pm 0.2$ & $1.8 \pm 0.1$ \\
M$_{\rm HI}$ (\msol) & $4.0 \times 10^5 D^2_{\rm Mpc}$  & $4.2 \times 10^5 D^2_{\rm Mpc}$ \\
M$_{\rm HI}$ (\msol) & $1.3\pm0.4 \times 10^7$  & $3.3\pm1.0 \times 10^7$ \\
M$_{\rm HI}$/L$_{\rm V}$ (\msol/L$_{\odot}$) & 4.6  &  3.8  \\
D$_{\rm HI}$ (10$^{19}$ cm$^{-2}$) & 11.7 kpc (7.2\arcmin) & 17.4 kpc (6.7\arcmin)  \\
D$_{\rm HI}$ ($3 \times 10^{18}$ cm$^{-2}$) & 15.6 kpc (9.6\arcmin) & 22.1 kpc (8.5\arcmin)  \\
V$_{\rm sys}$ (\kms) & $233 \pm 3$ &  $617 \pm 3$ \\
$\Delta V_{50}$ (\kms) & $28 \pm 3$ &  $41 \pm 3$ \\
V$_{\rm max}$ (\kms) & $4.6 \pm2.0$	& $9.4 \pm2.1$ \\
$\sigma$ (\kms) & $9.4\pm2.0$	 & $11.0\pm3.7$ \\
\hline
\end{tabular}
\end{table}


\section{Discussion}
\label{sec:dis}

What we can see from the derived velocity fields is that, while the disks are fairly regular on the approaching (blue) sides, they are strongly warped on the receding (red) sides. This asymmetry, combined with the strong gradient of velocity dispersion, shows clearly that those disks are not yet in equilibrium. 
An alternative scenario could be that some minor amount of accretion is happening in the outer regions, while the inner disks are fairly well in equilibrium.
Pisces A \& B are among the few non-elliptical (non-dSph) systems known  at $M_B \sim -11$, such as GR8 \citep{car90} in the Local Group and M81dwA  \citep{sar83} where, while rotation is providing some gravitational support in the inner parts, random motions provide essentially all the support in the outer parts. M81dwA has a rotation $\sim$3 \kms\, similar to Pisces A and GR8 has a rotation of $\sim$8 \kms\, of the same order as Pisces B.

When the KAT-7 \hi\ observations started, the distances to Pisces A \& B were unknown. Now that the distances are known, it is clear that those galaxies are not candidates for sub-halos of more massive MW-type galaxies. 
In the case of more likely sub-halo candidates, such as the 9 ultra faint dwarfs (UFD) uncovered by the Dark Energy Survey \citep{kop15}, none of them were detected in \hi\ \citep{wes15}.

Knowing the distances also revealed that Pisces A \& B are quite different from Leo T \& P, despite very similar HI fluxes. In fact, they are both much brighter with absolute V magnitude of $-11.57$ and $-12.90$, compared to $-7.1$ \citep{irw07} and $-9.3$ \citep{rho13} and have a much larger \hi\ content of a few $\times 10^7$ \msol, compared to a few $\times 10^5$ \msol\ for the less distant Leo T and Leo P.

\section{Conclusion}
\label{sec:con}

All isolated dwarfs (i.e. those that are not satellite of a larger galaxy such as the Milky Way or M31) are actually richer in cold gas than larger galaxies \citep{geh12} and, in fact, the smaller such isolated dwarfs are, the richer in cold gas they become \citep{hua12}. Some recent theoretical work \citep[see e.g.:][]{boa11, bob11} does indeed suggest that an extensive population of yet undiscovered gas-dominated, or even dark, dwarfs may exist in the outer fringes of the Local Group or just beyond. 

An examination of the dwarf galaxies within the full ALFALFA population in the context of
global star formation (SF) laws \citep{hua12} is consistent with the general assumptions that gas-rich galaxies have lower SF
efficiencies than do optically selected populations and that \hi\ disks are more extended than stellar ones. Pisces A \& B are good examples of this.

\begin{acknowledgements}

We thank all the team of SKA South Africa for allowing us to obtain scientific data during the commissioning phase of KAT-7. 
The work of CC an TJ is based upon research supported by the South African Research Chairs Initiative (SARChI) of the Department of Science 
and Technology (DST),  the Square Kilometre Array South Africa (SKA SA) and the National Research Foundation (NRF).
The research of YL, DL \& TR has been supported by SARChI, SKA SA fellowships.
\end{acknowledgements}



\begin{thebibliography}{}

\bibitem[Adams, Giovanelli \& Haynes(2013)]{ada13} Adams. E. A. K., Giovanelli, R. \& Haynes, M.P. 2013, \apj, 768, 77
\bibitem[Adams et al.(2015)]{ada15} Adams. E. A. K. et al 2015, \aap, 573, L3
\bibitem[Bovill \& Ricotti(2011a)]{boa11} Bovill, M. S.\& Ricotti, M. 2011a, \apj, 741, 17
\bibitem[Bovill \& Ricotti(2011b)]{bob11} Bovill, M. S.\& Ricotti, M. 2011b, \apj, 741, 18
\bibitem[Brooks et al.(2013)]{bro13} Brooks, A. M., Kuhlen, M., Molotov, A. \& Hooper, D. 2013, \apj, 765, 22
\bibitem[Cannon et al.(2015)]{can15} Cannon, J. M. et al. 2015,\aj, 149, 72
\bibitem[Carignan et al.(2013)]{car13} Carignan, C., Frank, B. S., Hess, K. M., Lucero, D. M., Randriamampandry, T. H., Goedhart, S. \& Passmoor, S. S. 2013, \aj, 146, 48
\bibitem[Carignan, Beaulieu \& Freeman(1990)]{car90} Carignan, C., Beaulieu, S. \& Freeman, K. C. 1990, \aj, 99, 1
\bibitem[Geha et al.(2012)]{geh12} Geha, M., Blanton, M. R., Yan, R. \& Tinker, J. L. 2012, \apj, 757, 85
\bibitem[Giovanelli et al.(2013)]{gio13} Giovanelli et al. 2013, \aj, 146, 15
\bibitem[Grcevich \& Putman(2003)]{grc09} Grcevich, J. \& Putman, M. E. 2009, \apj, 696, 385
\bibitem[Greisen(2003)]{gre03} Greisen, E. W., 2003,  Astrophysics and Space Science Library, 285, 109
\bibitem[Hess et al.(2015)]{hes15} Hess, K. M., Jarrett, T., Carignan, C., Passmoor, S. S., Goedhart, S. 2015, \mnras, 452, 1617
\bibitem[Huang et al.(2012)]{hua12} Huang, S., Haynes, M. P., Giovanelli, R., Brinchmann, J., Stierwalt, S. \& Neff, S. G. 2012, \aj, 143, 133
\bibitem[Irwin et al.(2007)]{irw07} Irwin, M. J. et al. 2007, \apj, 656, L13
\bibitem[Janowiecki et al.(2015)]{jan15} Janowiecki, S. et al. 2015, \apj, 801, 96
\bibitem[Koposov et al.(2015)]{kop15} Koposov, S. E., Belokurov, V., Torrealba, G. \& Evans, N. W. 2015, \apj, 805, 130
\bibitem[Laevens et al.(2015)]{lae15} Laevens, B. P. M. et al. 2015, arXiv:1507.07564
\bibitem[Lucero et al.(2015)]{luc15} Lucero, D. M., Carignan, C., Elson, E. C., Randriamampandry, T. H., Jarrett, T. H., Oosterloo, T. A. \& Heald, G. H. 2015, \mnras, 450, 3935
\bibitem[McConnachie(2012)]{mcc12} McConnachie, A. W. 2012, \aj, 144, 4
\bibitem[McMullin et al.(2007)]{mcm07} McMullin, J. P., Waters, B., Schiebel, D., Young, W. \& Golap, K. 2007, ASP Conference Series, 376, 127
\bibitem[McQuinn et al.(2013)]{mcq13} McQuinn, K. B. W. et al 2013, \aj, 146, 145
\bibitem[Merritt, van Dokkum \& Abraham(2014)]{mer14} Merritt, A., van Dokkum, P. \& Abraham, R. 2014, ApJ, 787, L37
\bibitem[Peek et al.(2011)]{pee11} Peek, J. E. G. et al. 2011, \apjs, 194, 20
\bibitem[Rhode et al.(2013)]{rho13} Rhode, K. L. et al. 2013, \aj, 145, 149
\bibitem[Sargent, Sancisi \& Lo(1983)]{sar83} Sargent, W. L. W., Sancisi, R. \& Lo, K.-Y. 1983, \apj, 265, 711
\bibitem[Spekkens et al.(2014)]{spe14} Spekkens, K., Urbancic, N., Mason, B., Willman, B. \& Aguirre, J. E. 2014, \apj, 795, L5
\bibitem[Tollerud et al.(2015)]{tol15} Tollerud, E. J., Geha, M. C., Grcevich, J., Putman, M. E. \& Stern, D. 2015, ApJ, 798, L21 (TOL15)
\bibitem[Tollerud et al.(2016)]{tol16} Tollerud, E. J., Geha, M. C., Grcevich, J., Putman, M. E., Weisz, D.  \& Dolphin, A. 2016 , submitted to \apj
\bibitem[Westmeier et al.(2015)]{wes15} Westmeier, T. et al. 2015, \mnras, 453, 338
\bibitem[Wong et al.(2006)]{won06} Wong, O.I. et al. 2006, \mnras, 371, 1855

\end{thebibliography}
\end{document}